\documentclass[fleqn,10pt]{wlscirep}
\usepackage[utf8]{inputenc}
\usepackage[T1]{fontenc}
\usepackage{epsfig}
\usepackage{amsfonts}
\usepackage{xcolor}
\usepackage{epsfig}
\usepackage{amsfonts}
\usepackage{xcolor}
\begin{document}

\newcommand{\bra}[1]{\left\langle#1\right|}
\newcommand{\ket}[1]{\left|#1\right\rangle}
\newcommand{\abs}[1]{\left|#1\right|}
\newcommand{\mean}[1]{\left\langle #1\right\rangle}
\newcommand{\braket}[2]{\left\langle{#1}|{#2}\right\rangle}
\newcommand{\commt}[2]{\left[{#1},{#2}\right]}
\newcommand{\tr}[1]{\mbox{Tr}{#1}}
	
\title{Symmetrized persistency of Bell correlations for Dicke states and GHZ-based mixtures: Studying the limits of monogamy}
\date{\today}		
\author[1,2,*]{Marcin Wie\'sniak}
\affil[1]{Institute of Theoretical Physics and Astrophysics, Faculty of Mathematics, Physics, and Informatics,\\ University of Gda\'nsk, 80-308 Gda\'nsk, Poland}
\affil[2]{International Centre for Theory of Quantum Technologies\\ University of Gda\'nsk, 80-308 Gda\'nsk, Poland}
\affil[*]{marcin.wiesniak@ug.edu.pl}
\begin{abstract}
Quantum correlations, in particular those, which enable to violate a Bell inequality \cite{BELL}, open a way to  advantage in certain communication tasks. However, the main difficulty in harnessing quantumness is its fragility to, e.g, noise or loss of particles. We study the persistency of Bell correlations of GHZ based mixtures and Dicke states. For the former, we consider quantum communication complexity reduction (QCCR) scheme, and propose new Bell inequalities (BIs), which can be used in that scheme for higher persistency in the limit of large number of particles $N$. In case of Dicke states, we show that persistency can reach $0.482N$, significantly more than reported in previous studies.  
\end{abstract}
\maketitle
\section{Introduction}
\subsection{State-of-Art and Motivation}
Harnessing quantum correlations can bring us unprecedented possibilities in communication and computation schemes. Core examples are quantum cryptographic key generation schemes \cite{BB84}, which allow an unbreakable message encryption. Their security is guaranteed by an impossibility of cloning of a quantum state, which follows directly from linearity of quantum mechanics. If Eve is trying to intercept a state sent between legitimate users, Alice and Bob, she must destroy quantum coherence, which draws the key generation impossible. In case of distributing entanglement, the same effect occurs by monogamy of entanglement, and in particular, monogamy of violation of CHSH inequality \cite{CHSH}. Simply put, one user (Alice) can violate inequality only with one other partner (Bob or Eve). Security of entanglement-based quantum cryptography was elegantly demonstrated in Ekert's E91 protocol \cite{EKERT91} .

It is then natural to extend the schemes of quantum cryptographic key distribution to more users, which collaborate to encrypt a message, so that it can be decrypted only by all of them. Such a scenario is called quantum secret sharing \cite{SECSHAR}, and its security was again linked with violation of multipartite BIs. It is known that, for example, violation of Werner-Wolf-Weinfurter-\.Zukowski-Brukner (WWW\.ZB) inequalities \cite{WW,WZ,ZB} is monogamous in a weak sense that only one of the inequalities among overlapping groups of observers can be violated maximally at the time, but not in a stronger sense, where only one can be violated at all. This potentially opens a loophole for an eavesdropper. A further generalization can be based on a subgroup of users, who must collaborate as a qualified majority to unlock a secret. 

A relevant protocol is quantum communication complexity reduction (QCCR) in distributed computing. Here, the task is to jointly compute a certain sign function under communication restrictions. Initially, it was shown that GHZ correlations \cite{GHZ} can reduce one bit of a necessary classical information exchange in certain situations \cite{ORIGCOMCOMP}.  Subsequently, it was reformulated as a quantum game, in which the goal is to optimize a guess of a 2$N$-bit function, where each user receives two bits and can broadcast only one. It has been shown that there is an advantage in this task once the function expresses a BI and the partners share a corresponding entangled state violating it \cite{COMCOMP}. 

Another concept that we consider here is persistency of quantum correlations. This quantity tells us how many parties must be traced out for a  given state to loose its property, e.g. entanglement (denoted as $(P^E(\rho))$, steerability $(P^S(\rho))$ (historically first Ref. \cite{BRIEGEL} gives a slightly different definition), or ability to violate a BI, somewhat misleadingly called ``nonlocality'' $(P^{NL}(\rho))$ \cite{VERTESI}. Hereafter, we will call the last kind persistency of Bell correlations and will be denoted as $P^{\text{Bell}}(\rho)$.

We will consider a stronger, symmetric version of persistency of Bell correlations, $P^{\text{Bell}}_{\text{sym}}(\rho)$. In other words, we will be interested in a number of observers (regardless of their identity) that need to be traced out in order for the observed statistics to have a local realistic description. We take two-fold context for this consideration. For mixtures based on GHZ states, we are mainly interested in the symmetrized quantum communication complexity reduction (sQCCR) game. We will ask what fraction of the total number of players, within an arbitrary ensemble, can achieve an advantage. Thus subsets of a certain cardinally can beat the classical game simultaneously. This consideration could be relevant in environments with significant particle losses or inefficient detectors.

In case of Dicke states \cite{WSTATE1}, we are more focused on fundamental aspects. In contrast to GHZ states, quantum correlations are present in all their reduced states of more than one qubit. Dicke states are hence strong candidates to show extremely high persistency of Bell correlations.  

\subsection{Monogamy of Bell correlations}
BIs distinguish between quantum-mechanical statistics and those, which permit a local realistic description. Thus, they are essential to recognize advantage Probably the simplest, but very useful one is the CHSH inequality. Consider two users, Alice and Bob, which have a pair of alternative observables, $A_1,A'_1$ for Alice, and $A_2,A'_2$ for Bob. Each of these observables can yield outcomes $+1$ or $-1$. The CHSH inequality then reads
\begin{eqnarray}
\mean{\textsl{B}_{12}}=&&\mean{A_1\otimes(A_2+A'_2)}+\mean{A'_1\otimes(A_2-A'_2)}\nonumber\\
\leq_{LR}&&2\nonumber\\,
\end{eqnarray}
where $\leq_{LR}$ means that the inequality holds for local realistic theories. The quantum mechanical mean value may reach $2\sqrt{2}\approx 2.82$. 

Now, consider a third observer, Eve, which also can has a choice of two local observables	$A_3,A'_3$. It has been shown in Ref. \cite{MONOGAMY} that when Bob performs a CHSH experiment simultaneously with Alice and Eve, only one of respective inequalities can be violated:
\begin{equation}
\mean{\textsl{B}_{12}}^2+\mean{\textsl{B}_{23}}^2\leq 8
\end{equation}
or in a weaker form \cite{OTHERMONOGAMY,OTHERMONOGAMY1}
\begin{equation}
\left|\mean{\textsl{B}_{12}}\right|+\left|\mean{\textsl{B}_{23}}\right|\leq 4
\end{equation}

 This result is crucial for the security of quantum cryptographic key distribution. When Eve entangles with legitimate users Alice and Bob, she must decrease quantum correlations between them. At the point Eve knows as much about the generated key as Alice and Bob, the CHSH inequality between the latter becomes satisfied.

The core of the proof lies in the necessary and sufficient condition for violating the inequality. Since we use only two observables per side, we can choose them to be strictly real, 
\begin{eqnarray}
A_i=&&(\sigma_x,\sigma_y,\sigma_z)\vec{c}_i\nonumber\\
A'_i=&&(\sigma_x,\sigma_y,\sigma_z)\vec{c}'_i\nonumber\\
\vec{c}_i=&&\left(\begin{array}{c}\cos\beta_i\\0\\ \sin\beta_i\end{array}\right	),\nonumber\\
\vec{c}'_i=&&\left(\begin{array}{c}-\sin\beta_i\\0\\ \cos\beta_i\end{array}\right	)
\end{eqnarray}
(hereafter, the second component is omitted). This choice of observables allows us to consider the reduced state to be strictly real, since the introduction of the nontrivial imaginary part would give the effect of state mixing. As observables of, say, Bob, are shared in both BIs, without a loss of generality strict realness can apply also to observable the third observer and reduced state between the Alice and Eve. 

Consider two observers. Notice that $A_1+A'_1=2\cos\beta_1(\sigma_x,\sigma_z)\vec{d}_1$ and $A_1-A'_1=2\sin\beta_1(\sigma_x,\sigma_z)\vec{d}'_1$, where $\vec{d}_1\perp\vec{d}'_1$ and $|\vec{d}_1|=|\vec{d}'_1|=1$. Thus, introducing $T_{ij}=\mean{\sigma_i\otimes\sigma_j}$, we get that 
\begin{eqnarray}
&&\mean{\textsl{B}_{12}}\nonumber\\
=&&2\left(\begin{array}{cc}T_{xx}&T_{xz}\\T_{zx}&TT_{zz}\end{array}\right)\left(\cos\alpha_2\vec{d_1}\otimes\vec{c_2}+\sin\alpha_2\vec{d'_1}\otimes\vec{c'_2}\right),
\end{eqnarray}
where $\vec{c}_2(\vec{c}'_2)=\cos\alpha_2\vec{d}_2+(-)\sin\alpha_2\vec{d}_2$. We now employ the Cauchy-Schwartz inequality, $\left|\vec{A}\cdot\vec{B}\right|\leq\sqrt{|\vec{A}|^2|\vec{B}|^2}$, where $\vec{A}=(T_{xx},T_{xz},T_{zx},T_{zz})^T$ and $\vec{B}=\cos\alpha_2\vec{d_1}\otimes\vec{c_2}+\sin\alpha_2\vec{d'_1}\otimes\vec{c'_2}$. Obviously $|\vec{B}|^2=1$, thus
\begin{equation}
\mean{\textsl{B}_{12}}^2\leq 4(T_{xx}^2+T_{xz}^2+T_{zx}^2+T_{zz}^2).
\end{equation}
In this case, the equality can be attained, as we have enough free parameters to conduct the Schmidt decomposition of the used sector of the correlation tensor. Going back to the three-user scenario we get 
\begin{eqnarray}
\label{mono1}
&&\frac{1}{4}\left(\mean{\textsl{B}_{12}}^2+\mean{\textsl{B}_{23}}^2\right)\nonumber\\
\leq &&T_{xx0}^2+T_{xz0}^2+T_{zx0}^2+T_{zz0}^2\nonumber\\
+ &&T_{0xx}^2+T_{0xz}^2+T_{0zx}^2+T_{0zz}^2,
\end{eqnarray}
where $T_{xx0}=\mean{\sigma_x\otimes\sigma_x\otimes\sigma_0}$, etc. Let us now use the methods presented in Refs. \cite{KPRLK,WIESMARU}. We create a graph with 8 vertices associated with the operators, means of which enter Ineq. (\ref{mono1}), and connect them if they anticommute. We get a cuboid, in which two opposite face have vertices connected on diagonals, as depicted in Figure 1. Next, we assign 0s and 1s to the vertices in such a way that no pair of 1s can be connected with an edge. Assigning 1 to any vertex eliminates four connected with it, and the remaing three are in a clique, so only one other 1 can be distributed among them. We thus get
\begin{eqnarray}
\label{mono2}
&&T_{xx0}^2+T_{xz0}^2+T_{zx0}^2+T_{zz0}^2\nonumber\\
+&&T_{0xx}^2+T_{0xz}^2+T_{0zx}^2+T_{0zz}^2\nonumber\\
\leq&&2,
\end{eqnarray}
meaning that once $\left|\mean{\textsl{B}_{12}}\right|$ goes above 2, $\left|\mean{\textsl{B}_{23}}\right|\leq 2$ cannot be violated and vice versa. In  this fashion, we can investigate if the strong monogamy relations hold for other inequalities, in particular, WWW\.ZB BIs. A rule of thumb is that if all subsets of observers have non-zero overlap, the bound of the sum is $2^{N_0}$, $N_0$ being the cardinality of the largest of these subsets. This happens when parties from this subset share a GHZ state, and hence other parties must be uncorrelated.

Now, consider the case of five parties labelled as A, B, C, D, and E. A, and E measure $\frac{1}{\sqrt{2}}(\sigma_x\pm\sigma_y)$, while B, C, and D measure $\sigma_x$ or $\sigma_y$. In one half of the runs, A,B,C, and D receive a GHZ state, and E receives the white noise, in the other half the roles of A and E are interchanged. If the first four observers shared a pure GHZ state, they would get the violation of a Mermin-Ardehali-Belinskii-Klyshko (MAKB) inequality \cite{MERMIN,ARDEHALI,BK} by factor $2\sqrt{2}$, but the state has effectively $50\%$ of noise. Thus, both ensembles, $\{$A, B, C, D$\}$ and  $\{$B, C, D, E$\}$, can simultaneously violate a MAKB inequality. Hence we have 
\begin{eqnarray}
&&P^{\text{Bell}}\left(1/4\left(\ket{GHZ_4}\bra{GHZ_4}\otimes\mathbb{1}_{2\times 2}\right.\right.\nonumber\\&&\left.\left.+\mathbb{1}
_{2\times 2}\otimes\ket{GHZ_4}\bra{GHZ_4}\right)\right)\nonumber\\
&=&2,
\end{eqnarray}
where
\begin{eqnarray}
|GHZ_l\rangle=&&\frac{1}{\sqrt{2}}\left(\ket{0}^{\otimes l}+\ket{1}^{\otimes l}\right).
\end{eqnarray}
I general we shall use ``greater than'' sign, rather than ``equal to'' for persistency of Bell correlations, as we cannot claim that we use the best inequalities. In this case, we clearly use the optimal inequality.

We will investigate a symmetrization of this scenario.

\subsection{Quantum Communication Complexity Reduction}
Here we briefly recall the idea behind the quantum advantage in communication complexity reduction problems in distributed computing related to BIs. This link was established in Ref. \cite{COMCOMP}, but it is only one-way \cite{NOCOMCOMP}. Originally, it was shown that when using a GHZ state, a certain function can be computed if users exchange one bit less in total \cite{ORIGCOMCOMP}, but Ref. \cite{COMCOMP} introduced the following probabilistic interpretation. Imagine $N$ users, whose task is to jointly calculate a certian dichotomic function. Each user receives two random variables from a dealer: a random bit $y_i=\pm 1$, with promise of  $P(y_i=+1)=1/2$, and $x_i$, which can be from any set, and the joint distribution of $x_i$s is promised,
\begin{equation}
P(x_1,...,x_N)=\frac{|g(x_1,...,x_N)|}{\sum_{x_1,...,x_N}|g(x_1,...,x_N)|}
\end{equation}
(or with integrals in the denominators). Now, a user can  perform an arbitrary local action, but must return (broadcast) one bit. From these bits they guess the value of function
\begin{equation}
F(x_1,...,x_N,y_1,...,y_N)=y_1...y_N\frac{g(x_1,...,x_N)}{|g(x_1,...,x_N)|}=\pm 1.
\end{equation}
The ultimate task is to yield the correct value of $F(x_1,...,x_N,y_1,...,y_N)$ in as many cases as possible.

Obviously, broadcasted bits must be contain information about $y_i$s, since omitting any one of them completely  destroy the correlation between the actual and the anticipated value. Then, if all $g(x_1,...,x_N)$s are of the same sign,  or 0, the task trivializes. If the sign varies, in the classical case the users are limited to broadcast $y_if_i(x_i)=\pm 1$. Thus they are restricted to local deterministic predictions. When entangled state $\ket{\Psi}$ of $N$ qubits is distributed among them, they can, however, make a measurement on a qubit they hold dependent on $x_i$,  obtain result $m_i$, and broadcast $y_im_i$. Then, if $g(x_1,...,x_N)$s are coefficient of a BI violated by $\ket{\Psi}$, the users take benefit from quantum correlations, and get a more efficient estimation of $F(x_1,...,x_N,y_1,...,y_N)$. Thus, this variant is will be hereafter QCCR game. 

We will be interested in a bit modified variant of this scheme, the sQCCR game.  We will still have $N$ users, but we demand that only $k$ is trying to estimate the function. Our restriction, though, is that this could be any subset of $k$ users, or, equivalently, each such group tries to estimate the function independently. For the sQCCR game, we additionally require that the marginal probabilities are symmetric under permutations of parties, i.e.,
\begin{equation}
\label{QCCR}
P(x_{\pi_1},x_{\pi_2},...,x_{\pi_k})=\sum_{x_{\pi_{k+1}},...,x_{\pi_N}}P(x_{\pi_1},x_{\pi_2},...,x_{\pi_N}),
\end{equation}
where $(\pi_1,...,\pi_N)$ is an arbitrary permutation of $(1,...,N)$.

We will also consider violation of a BI under such restrictions. The difference between a mere BI violation and QCCR scheme lies in the demand that in latter case, the measurements settings are distributed with a known probability distribution. It might be impossible to find a distribution that has a desired one as all marginals of $k$th order. In case of $BI$s, user draw the measurements settings locally and independently to close the common cause loophole, so frequency of their appearance only affects the trust level of average values. Most optimally, a flat distribution is used.
\section{Results}
\subsection{Symmetrized persistency of Bell correlations for GHZ-based mixtures}
First, let us discuss the case of GHZ states. Obviously, any quantum advantage for $k<N$ of $N$ users is not possible for pure states, as any reduced state is fully separable. We thus need to use a symmetrized mixture,
\begin{equation}
\rho=\frac{1}{2^{N-k}N!}\sum_{\Pi}\Pi\left(\ket{GHZ_k}\bra{GHZ_k}\otimes\mathbb{1}_{2^{N-k}\times 2^{N-k}}\right),
\end{equation}
where	$\sum_{\Pi}Pi(\cdot)$ denotes the sum over all permutations of parties. We refer to these states as GHZ based mixtures.

Let us start with MAKB inequalities, which are obtained in an iterative way. Each observer has choice two observables, $A_i$ and $A'_i$. The Bell expressions are
\begin{eqnarray}
&&B_2(A_1,A'_1,A_2,A'_2)\nonumber\\
=&&\frac{1}{2}(A_1\otimes(A_2+A'_2)+A'_1\otimes(A_2-A'_2)),\nonumber\\
&&B_{i+1}(A_1,A'_1,...,A_{i+1},A'_{i+1})\nonumber\\
=&&\frac{1}{2}((A_{i+1}+A'_{i+1})\otimes B_i(A_1,A'_1,....,A_i,A'_i)\nonumber\\
+&&(A_{i+1}-A'_{i+1})\otimes B_i(A'_1,A_1,....,A'_i,A_i)
\end{eqnarray}
The maximal local realistic values are $2^{\frac{N-1}{2}}$ for odd $N$ and $2^{\frac{N}{2}}$ for even. If we take $A_i=\cos(2\pi\alpha_i) \sigma_x+\sin(2\pi\alpha_i) \sigma_y$ and likewise for $A'_i$s, we get
\begin{equation}
\bra{GHZ_N}A_1\otimes...\otimes A_N\ket{GHZ_N}=\cos\left(2\pi\sum_i\alpha_i\right).
\end{equation} 
For odd $N$ we have $2^{N-1}$ average values, and with choice of observables $A_i=\sigma_x$, $A'_i=\sigma_y$ all of them have modulo 1 and a sign corresponding to the respective terms in the Bell expression. Thus the maximal quantum mechanical value is $2^{N-1}$, and the quantum-to-classical ratio (QCR), the ratio between the maximal quantum and classical values, is $2^{(N-1)/2}$. For even $N$, again, all signs are matched, but the optimal modulo is $1/\sqrt{2}$ an QCR is still $2^{(N-1)/2}$. A symmetrized optimal choice of observables is $\alpha_i=\frac{1}{8N}$ and $\alpha'_i=\frac{2N+1}{8N}$. 

In case of GHZ states, even stronger inequalities were found. They utilize a continuum of local observables, i.e. $\alpha_i$ will take an arbitrary value. Naturally, any feasible implementation of these inequalities will utilize a finite number of uniformly distributed vales of $\alpha_i$ \cite{FINITEA}. The coefficiens are given by the values of the quantum correlation function and the QCR is $\frac{1}{2}\left(\frac{\pi}{2}\right)^N$

\subsection{Symmetrized persitency of Bell correlations in QCCR protocols}
As we have seen, for GHZ-based mixtures, QCR grows exponentially with the number of parties, while the symmetrization causes only a polynomial  decay of correlation. Thus for any value $N-k$ there is some value of $N$, above which any $N-k$ users can be traced out. 

For MAKB inequalities and geometrical BIs (GBIs) \cite{GBI0,GBI} QCR and $M$-partite GHZ states behaves like 
\begin{equation}
QCR(N)=b\times a^M,
\end{equation} 
where $a=\sqrt{2}$, $b=1/\sqrt{2}$ for MAKB inequalities and $a=\pi/2$, $b=1/2$ for GBIs. Let us now extend and symmetrize the state so that any $N-M$ users can violate a BI,
\begin{equation}
\left(\begin{array}{c}N\\M\end{array}\right)^{-1}b\times a^M>1,
\end{equation}
which is the condition for $P^{\text{Bell}}_{\text{sym}}>M$. $P^{\text{Bell}}_{\text{sym}}\geq 2$ occurs for $N=9$ in case of MAKB inequalities and $N=7$ for GBIs. Let us now study the asymptotic behavior with $N\rightarrow\infty$ by referring to the Stirling approximation and introducing $\gamma=M/N$.
\begin{eqnarray}
0<&&H(\gamma)-\gamma \log_2{a}-\frac{\log_2{b}}{N},\nonumber\\
H(x)=&&-x \log_2{x}-(1-x)\log_2(1-x).
\end{eqnarray}
Note that $\frac{\log_2{b}}{N}$ is neglible for large $N$. We are thus left with 
\begin{equation}
\label{EQ20}
H(\gamma)>\gamma \log_2{a}.
\end{equation}
We can now find the ratio of parties that must be preserved. This cannot be done analytically, but is guaranteed to happen for any $a$: $H(0)=0$,$\left.\partial H(x)/\partial x\right|_{x=0}=+\infty$, $H(1-x)=H(x)$, and $H(x)\leq 1$. For $a=\sqrt{2}$ Ineq. (\ref{EQ20}) is satisfied for $\gamma<\gamma_{CRIT}=0.905118$, whereas $a=\pi/2$ gives $\gamma_{CRIT}=0.867227$. 

Thus, in the limit of large $N$ up to $13.2\%$ of users can be replaced by others in the protocol. However, gaining an advantage in communication complexity reduction scheme requires an extra discussion. As we have mentioned above, in case of a mere Bell test distribution of measurement settings is largely irrelevant, and it is even desired to be uniform, in the QCCR scheme this distribution encodes the BI. For example, for an even number of parties with MAKB inequalities, we use all combinations of observables with equal weigths, but only a half of them for odd. To enjoy the benefit in QCCR  we need to have even $k$. Otherwise, after replacing one of the partners with another one we would not be able to recreate the distribution. 

A similar problem arises with GBIs. However, this can be easily fixed by introducing new GBIs, in which observables from the $x-y$ plane (with eigenvalues $\pm 1$). Thus the quantum mechanical part reads
\begin{eqnarray}
Q_N=&&\int_0^1d\alpha_1...\int_0^1d\alpha_N\nonumber\\
\times&&\text{sign}\left(\cos\left(2\pi\sum_i\alpha_i\right)\right)\cos\left(2\pi\sum_i\alpha_i\right)\nonumber\\
=&&\int_0^1d\alpha_1...\int_0^1 d\alpha_N\left|\cos\left(2\pi\sum_i\alpha_i\right)\right|\nonumber\\
=&&\frac{2}{\pi}.
\end{eqnarray}
The optimal classical part is equal to
\begin{eqnarray}
C_N=&&2^N\int_{-N/4}^{(-N+1)/4}d\alpha_1\int_{0}^{1/2}d\alpha_2...\int_{0}^{1/2}d\alpha_N\nonumber\\
\times&&\text{sign}\left(\cos\left(2\pi\sum_i\alpha_i\right)\right)
\end{eqnarray}
and $\{C_2,\,C_3,\,C_4,\,C_5,\,C_6,\,C_7,...\}\,=\{1/2,\,1/3,\,5/24,\,2/15,\,61/720,\,17/315\,...\}$. We have 
\begin{equation}
\lim_{N\rightarrow\infty}\frac{C_{N-1}}{C_N}=\frac{\pi}{2}.
\end{equation}
and the convergence is exponential. The first instance of persistency of $P^{\text{Bell}}_{\text{sym}}\geq2$ occurs for $N=7$, ${Q_6}/(7{C_6})=1440/(427\pi)\approx 1.07346$.

\subsection{Symmetrized persistancy of Bell correlations for Dicke states}
Dicke state family is exemplary for studying persistency of Bell correlations,
\begin{eqnarray}
\ket{D_{N,M}}=&&\left(\begin{array}{c}N\\M\end{array}\right)^{-\frac{1}{2}}\Pi(|0\rangle^{\otimes M} |1\rangle^{\otimes N-M}),
\end{eqnarray}
where $\Pi(\cdot)$ denotes the sum over all permutations of parties. In particular, Refs. \cite{WSTATE1,WSTATE2,WSTATE3,WSTATE4,VERTE} studied $P^{\text{Bell}}_{\text{sym}}$ for W states ($\ket{W_N}=\ket{D_{N,1}}$). In particular, the Authors of Ref. \cite{VERTE} considered BIs involving correlation between subsets of observers. Another important contribution of Ref. \cite{VERTE} was an observation that un upper bound for $P^{\text{Bell}}_{\text{sym}}(\ket{W_N}\bra{W_N})$ is $N/2$. With the inequalities found therein, they found $P^{\text{Bell}}_{\text{sym}}(\ket{W_N}\bra{W_N})\geq 2/5N$. In this work we limit ourselves to WWW\.ZB inequalities. The respective Bell operators read
\begin{eqnarray}
\label{Eq16}
B'_N=&&\frac{1}{2^N}\sum_{s_1,...,s_N=\pm 1}S(s_1,...,s_N)\nonumber\\
\times&&(A_1+s_1A'_1)\otimes...\otimes(A_N+s_NA'_N)\nonumber\\
\leq&& 1,\nonumber\\
S(s_1,...,s_N)=&&\pm 1.
\end{eqnarray}
We thus have a choice of $2^{2^N}$ different operators, one for each sign function $S(s_1,...,s_N)$ (naturally, they can be either trivial, or trivially related amongst them).

The necessary condition for violation of WWW\.ZB inequalities is that there exists such a choice of local directions $x$ and $z$, that 
\begin{equation}
\label{viol}
\sum_{i_1=x,z}...\sum_{i_N=x,z}T^2_{i_1...i_N}>1.
\end{equation}

The problem with violation of WWW\.ZB inequalities with Dicke states is that there is a violation gap. That is to say, 	there is a certain range of white noise admixture, for which the state satisfies condition 	(\ref{viol}), but still does not violate any inequality from this family. However, this gap is relatively small (typically less than $1\%$ of the sum in Ineq. (\ref{viol})), hence the sum is hence a good and fast 	indicator of potency for violation.  Additionally, the same condition becomes necessary and sufficient for violation of inequalities with more settings per side under the restriction that the observables lie in the $x-z$ plane \cite{MANY}.

The partial trace of the Dicke state over $L$ parties is 
\begin{eqnarray}
&&\rho_{N,M,L}\nonumber\\
=&&\text{Tr}_L(\ket{D_{N,M}}\bra{D_{N,M}})\nonumber\\
=&&\left(\begin{array}{c}N\\M\end{array}\right)^{-1}\sum_{l=0}^L\left(\begin{array}{c}L\\l\end{array}\right)\left(\begin{array}{c}N-L\\M-l\end{array}\right)\nonumber\\
\times&&\ket{D_{N-L,M-l}}\bra{D_{N-L,M-l}}.
\end{eqnarray}
	
In the next step we calculate 
\begin{eqnarray}
&&\sum_{i_1=x,z}...\sum_{i_{N-L}=x,z}T^2_{i_1...i_{N-L}}(N,M,L)\nonumber\\
&=&\sum_{i_1=x,z}...\sum_{i_{N-L}=x,z}\text{Tr}\rho_{N,M,L}(\sigma_{i_1}\otimes...\otimes\sigma_{i_{N-L}}).
\end{eqnarray}
These data are then interpolated to function $\Sigma_{M,L}(N)$ and equation $\Sigma_{M,L}(N_0)=1$ is solved for $N_0$. 
The values of $N_0$ are given in table 1.
If the results hold the pattern for larger $M$, we shall have $a\approx 2.0925	1+1/M$.  
We thus have $P^{\text{Bell}}_{\text{sym}}\geq 2$ for $(N,M)\in\{(5,1),(6,2),(8,3),(9,4)\}$. As we can see, in	 the limit of both large $N$ and large $L$ we estimate that $P^{\text{Bell}}_{sym}\geq 0.482N$

In case of W states we have studied some particular inequalities. None of them  satisfied condition (\ref{QCCR}). Thus we assume that they are not useful for the sQCCR game.   
\section{Conclusions}
We have studied symmetrized Bell correlations persistency for mixtures of GHZ states and Dicke states, especially in the context of their link to quantum communication schemes. We found that mixtures based on GHZ states asymptotically allow for loss of about $9.5\%$ to MAKB inequalities and $13.2\%$ for GBIs. 

We have characterized the persistency of Bell correlations of Dicke states with respect to WWW\.ZB inequalities. We have observed that already for three excitations we have asymptotic persitency higher than $2/5N$, and in the limit of large number of excitations, it shall reach $0.477N$. Thus, contrary to the remarks in Ref. \cite{VERTE}, many-excitation Dicke states perform significantly better than the reported persitency for W states, even though we have not tailored BIs for the former. 

We have found that in a symmetrized situation described in the current paper users sharing the GHZ states-based mixture can still enjoy the quantum advantage in the sQCCR game in subgroups of up to $0.89N$ users. They can achieve this using even-number-of-parties MAKB BIs, or a variant of GBIs, which asymptotically has quantum-to-classical ratio proportional to $(\pi/2)$   
\section*{Acknowledgements}
h This work is a part of NCN grant No. UMO-2017/26/E/ST2/01008. The Author acknowledges the ICTQT IRAP project of FNP, financed by structural funds of EU.
\section*{Author Contribution Statement}
MW has conceptualized and conducted the research, written and reviewed the manuscript.
\section*{Statement of a conflict of interests}
The Author declares no conflict of interests.

\begin{figure}[!ht]
\label{Fig1}
\includegraphics[width=8cm]{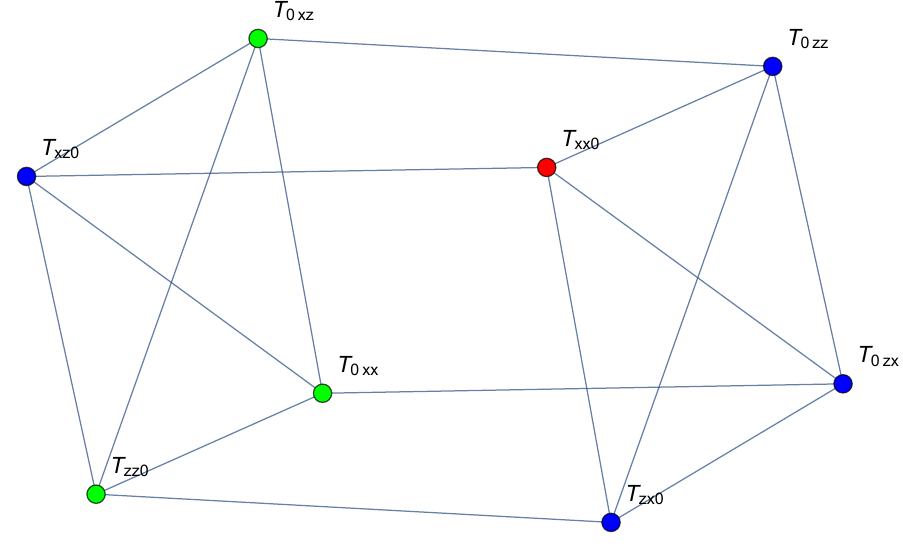}
\caption{(Color online) Anticommutativity graph for two CHSH inequalities with one common observer. We can assign value ``1'' to one of the vertices (red), which forces four other to take value ``0'' (blue). This will leave three vertices with unassigned value (green)}
\end{figure}

\begin{table}[!ht]
\begin{tabular}{|c|c|c|}
\hline
M&a&b\\
\hline					
1&3&1\\
2&2.5776&2.8083\\
3&2.4043&3.6349\\
4&2.3325&6.4408\\
\hline
\end{tabular}
\caption{Values for estimating $N_0=a L+b$, above which a WWW\.ZB inequality can be violated. $1/a$ is a ratio between asymptotic $P^{\text{Bell}_{sym}}$ and $N$.}
\end{table}

\end{document}